\documentclass[12pt]{article}

\usepackage{amsmath}
\usepackage{amssymb}
\usepackage[dvips]{graphicx}
\usepackage{cite}

\makeatletter
 
  \@addtoreset{equation}{section}
 \makeatother

\newcommand {\n}{\nonumber \\}
\newcommand {\tr}{\mbox{tr}}
\newcommand {\Tr}{\mbox{Tr}}

\begin{document}
\setlength{\oddsidemargin}{0cm}
\setlength{\baselineskip}{7mm}

\begin{titlepage}
\begin{normalsize}
\begin{flushright}
\begin{tabular}{l}
March 2010
\end{tabular}
\end{flushright}
  \end{normalsize}

~~\\

\vspace*{0cm}
    \begin{Large}
       \begin{center}
         {Model of M-theory with Eleven Matrices}
       \end{center}
    \end{Large}
\vspace{1cm}

\begin{center}
           Matsuo S{\sc ato}\footnote
           {
e-mail address : msato@cc.hirosaki-u.ac.jp}\\
      \vspace{1cm}
       
         {\it Department of Natural Science, Faculty of Education, Hirosaki University\\ 
 Bunkyo-cho 1, Hirosaki, Aomori 036-8560, Japan}

\end{center}

\hspace{5cm}

\begin{abstract}
\noindent
We show that an action of a supermembrane in an eleven-dimensional spacetime with a semi-light-cone gauge can be written only with Nambu-Poisson bracket and an invariant symmetric bilinear form under an approximation. Thus, the action under the conditions is manifestly covariant under volume preserving diffeomorphism even when the world-volume metric is flat. Next, we propose two 3-algebraic models of M-theory which are obtained as a second quantization of an action that is equivalent to the supermembrane action under the approximation. The second quantization is defined by replacing Nambu-Poisson bracket with finite-dimensional 3-algebras' brackets. Our models include eleven matrices corresponding to all the eleven space-time coordinates in M-theory although they possess not $SO(1,10)$ but $SO(1,2) \times SO(8)$ or $SO(1,2) \times SU(4) \times U(1)$ covariance. They possess $\mathcal{N}=1$ space-time supersymmetry in eleven dimensions that consists of 16 kinematical and 16 dynamical ones. We also show that the $SU(4)$ model with a certain algebra reduces to BFSS matrix theory if DLCQ limit is taken.

\end{abstract}

\vfill
\end{titlepage}
\vfil\eject

\setcounter{footnote}{0}

\section{Introduction}
\setcounter{equation}{0}

BFSS matrix theory is conjectured to describe infinite momentum frame (IMF) limit of M-theory in \cite{BFSS} and many evidences were found. However, because of the limit, the theory does not include a variable corresponding to the eleventh space-time coordinate of M-theory; it includes only time and nine matrices corresponding to nine spatial coordinates. As a result, it is very difficult to derive full dynamics of M-theory. For example, we do not know the manner to describe longitudinal momentum transfer of D0-branes. 
Therefore, we need a model that includes variables corresponding to all the eleven space-time coordinates in M-theory.

Recently, dynamics of supermembranes in M-theory have been investigated intensively. New superconformal field theories in three dimensions were constructed in \cite{BLG1, Gustavsson, BLG2} and the low energy effective action of N multiple supermembranes was found in \cite{ABJM, N=6BL}. Because the effective action possesses a symmetry based on a 3 algebra,
3 algebras are expected to play a crucial role in constructing a model of M-theory \cite{Iso, talk, bosonicM, LeePark1, BUTSURI}.

The supermembrane action in light-cone gauge and Green-Schwarz type IIB superstring action in Schild gauge \cite{deWHN, IKKT} can be written only with Poisson bracket and an invariant symmetric bilinear form. As a second quantization, by replacing the Poisson bracket in the actions with a Lie-bracket, we obtain BFSS matrix theory and type IIB matrix model, respectively. While the original membrane and  string actions describe a single object, the matrix models describe many body interactions. Although the IIB matrix model is a covariant constructive formulation of type IIB superstring theory, BFSS matrix theory is not covariant because of the light-cone gauge. On the other hand, the bosonic part of a supermembrane action can be written in a $SO(1, 10)$ covariant form as $T_{M2} \int d^3 \sigma \sqrt{g}\left(-\frac{1}{12}(\frac{1}{\sqrt{g}}\{ X^L, X^M, X^N \})^2+\Lambda\right)$, where $L, M, N = 0, \cdots, 10$ and $\{\, , \, , \, \}$ is Nambu-Poisson bracket \cite{Nambu, Yoneya}. Although we can obtain a covariant second quantized bosonic model by replacing Nambu-Poisson bracket with a finite-dimensional 3-algebra's bracket \cite{bosonicM}\footnote
{
A formulation of M-theory by a cubic matrix action was proposed by Smolin \cite{Smolin1, Smolin2, Azuma} }, we cannot obtain a full action of the second quantized model because it seems impossible to write the full supermembrane action only with Nambu-Poisson bracket and an invariant symmetric bilinear form covariantly. For example, see \cite{LeePark2}.

In section two of this paper, we show that a supermembrane action with a semi-light-cone gauge
 can be written only with Nambu-Poisson bracket and an invariant bilinear symmetric form under an approximation. Thus, the action under the conditions is manifestly covariant under volume preserving diffeomorphism (VPD) even when the world-volume metric is flat. In section three, we propose two 3-algebraic models of M-theory which are obtained as a second quantization of an action that is equivalent to the supermembrane action under the approximation. The second quantization is defined by replacing Nambu-Poisson bracket with finite-dimensional 3-algebras' ones.  Our models include matrices corresponding to all the eleven space-time coordinates in M-theory because they possess the same structure as the supermembrane action with a semi-light-cone gauge, where only the kappa symmetry is fixed and bosonic space-time coordinates are not constrained\footnote
{
Advantages of a semi-light-cone gauges against a light-cone gauge are shown in \cite{Carlip, Kallosh, KazamaYokoi}}. Because the 32 fermions are constrained to be 16 fermions, the models possess only 16 kinematical supersymmetries. These and 16 dynamical supersymmetries of the models form $\mathcal{N}=1$ space-time supersymmetry of M-theory as in \cite{MSUSY}. In section four, we show that the $SU(4)$ model with a certain algebra reduces to BFSS matrix theory if DLCQ limit is taken.

\vspace{1cm}

\section{VPD Covariance of Supermembrane Action}
\setcounter{equation}{0}
We start with an action,
\begin{eqnarray}
S_{cl}
&=&
\int d^3 \sigma \sqrt{-g}
\Bigl(
-\frac{1}{12}\{X^I, X^J, X^K\}^2 
-\frac{1}{2}(A_{\mu a b} \{\varphi^a, \varphi^b, X^{I}\})^2  \n
&& \qquad \qquad
-\frac{1}{3} E^{\mu \nu \lambda} 
A_{\mu a b} A_{\nu c d} A_{\lambda e f} 
\{\varphi^a, \varphi^c, \varphi^d\}\{\varphi^b, \varphi^e, \varphi^f\} 
+\frac{1}{2}\Lambda  \n
&& \qquad \qquad
-\frac{i}{2}\bar{\Psi} \Gamma^{\mu} A_{\mu a b} \{\varphi^a, \varphi^b, \psi\}
+\frac{i}{4}\bar{\psi}\Gamma_{IJ}\{X^I, X^J, \psi\} 
\Bigr), 
\label{continuumaction}
\end{eqnarray}
where $I, J, K = 3, \cdots, 10$ and $\{ \varphi^a, \varphi^b, \varphi^c \}=\epsilon^{\alpha \beta \gamma} \partial_{\alpha} \varphi^a \partial_{\beta} \varphi^b \partial_{\gamma} \varphi^c$ $(\alpha, \beta, \gamma = 0, 1, 2)$ is Nambu-Poisson bracket. An invariant symmetric bilinear form is defined by $\int d^3 \sigma \sqrt{-g} \varphi^a \varphi^b$ for complete basis $\varphi^a$ in three dimensions. Thus, this action is manifestly VPD covariant even when the world-volume metric is flat. 
$X^I$ is a scalar and $\psi$ is a $SO(1,2) \times SO(8)$ Majorana-Weyl fermion satisfying
\begin{eqnarray}
&&\Gamma^{012}\psi=-\psi \label{kappa} \\
&& \psi^{\dagger} = \psi^{T}.
\end{eqnarray}
$E^{\mu \nu \lambda}$ is a Levi-Civita symbol in three dimensions and $\Lambda$ is a cosmological constant. We will show that this action is equivalent to a supermembrane action with a semi-light-cone gauge (\ref{kappa}) under an approximation.

$A_{\mu a b}$ $(\mu = 0, 1, 2)$ one-to-one corresponds to a bi-local field $A_{\mu}(\sigma, \sigma')$ by $A_{\mu}(\sigma, \sigma')=A_{\mu a b}\varphi^a(\sigma)\varphi^b(\sigma')$ and $A_{\mu}(\sigma, \sigma')$ consists of infinite numbers of local tensors as \cite{M5fromM2, M5withC}
\begin{equation}
A_{\mu}(\sigma, \sigma')
=\partial'_{\alpha} A_{\mu}(\sigma, \sigma')|_{\sigma'=\sigma} (\sigma^{'\alpha}-\sigma^{\alpha})
+\frac{1}{2}\partial'_{\alpha}\partial'_{\beta}A_{\mu}(\sigma, \sigma')|_{\sigma'=\sigma}(\sigma^{' \alpha}-\sigma^{\alpha})(\sigma^{' \beta}-\sigma^{\beta})+ \cdots.\label{infinitemodes}
\end{equation}
However, $A_{\mu a b}$ appears only in a combination $A_{\mu a b} \partial_{\alpha} \varphi^a \partial_{\beta} \varphi^b$ in (\ref{continuumaction}), namely among infinite numbers of modes in (\ref{infinitemodes}) only the vector mode 
\begin{equation}
a_{\mu \alpha} \equiv \partial'_{\alpha} A_{\mu}(\sigma, \sigma')|_{\sigma'=\sigma}=A_{\mu a b} \varphi^a \partial_{\alpha} \varphi^b,
\end{equation}
contributes to the action. The combination is rewritten as
\begin{equation}
A_{\mu a b} \partial_{\alpha} \varphi^a \partial_{\beta} \varphi^b
=
\frac{1}{2}(\partial_{\alpha} a_{\mu \beta}-\partial_{\beta} a_{\mu \alpha})
=
\frac{1}{2}F_{\mu \alpha \beta}. \label{combination}
\end{equation}
By using this, the action is given by
\begin{eqnarray}
S_{cl}
&=&
\int d^3 \sigma \sqrt{-g}
\Bigl(
-\frac{1}{12}(\epsilon^{\alpha \beta \gamma} \partial_{\alpha} X^I \partial_{\beta} X^J \partial{\gamma} X^K)^2
-\frac{1}{8}(\epsilon^{\alpha \beta \gamma} F_{\mu \alpha \beta} \partial_{\gamma} X^I)^2\n
&& \qquad \qquad \qquad 
-\frac{1}{6} E^{\mu \nu \lambda} \epsilon^{\alpha \beta \gamma} \epsilon^{\alpha' \beta' \gamma'} F_{\mu \alpha \alpha'} F_{\nu \beta \gamma} F_{\lambda \beta' \gamma'}) + \frac{1}{2}\Lambda \n
&& \qquad \qquad \qquad 
-\frac{i}{4} \bar{\psi} \Gamma^{\mu} \epsilon^{\alpha \beta \gamma} F_{\mu \alpha \beta} \partial_{\gamma} \psi 
+\frac{i}{4} \bar{\psi} \Gamma_{IJ} \epsilon^{\alpha \beta \gamma} \partial_{\alpha} X^I \partial_{\beta} X^J \partial_{\gamma} \psi
\Bigr).
\end{eqnarray}
We can regard $F_{\mu \alpha \beta}$ as a fundamental field by introducing auxiliary field $X^{\mu}$ and by adding a term
\begin{equation}
\int d^3 \sigma \sqrt{-g}
\Bigl(-\frac{1}{2}X^{\mu} \epsilon^{\alpha \beta \gamma} \partial_{\alpha} F_{\mu \beta \gamma}
\Bigr)
\end{equation}
to the action. Next, we take Poincare dual of $F_{\mu \alpha \beta}$ in three dimensions,
\begin{equation}
F_{\mu \alpha \beta}= \epsilon_{\alpha \beta \gamma} A_{\mu}^{\,\,\, \gamma}. \label{PoincareDual}
\end{equation}
Thus, we obtain
\begin{eqnarray}
S_{cl}
&=&
\int d^3 \sigma \sqrt{-g}
\Bigl(
-\frac{1}{12}(\epsilon^{\alpha \beta \gamma} \partial_{\alpha} X^I \partial_{\beta} X^J \partial{\gamma} X^K)^2
-\frac{1}{2}(A_{\mu}^{\,\,\,\alpha}\partial_{\alpha} X^I)^2 \n
&& \qquad \qquad
-\frac{1}{6} E^{\mu \nu \lambda} 
\epsilon^{\alpha \beta \gamma} 
A_{\mu \alpha} A_{\nu \beta} A_{\lambda \gamma}
+ \frac{1}{2}\Lambda 
-\partial_{\alpha} X^{\mu} A_{\mu}^{\,\,\,\alpha} 
\n
&& \qquad \qquad
+\frac{i}{2} \bar{\psi} \Gamma^{\mu} A_{\mu}^{\,\,\,\alpha} \partial_{\alpha} \psi 
+\frac{i}{4} \bar{\psi} \Gamma_{IJ} \epsilon^{\alpha \beta \gamma} \partial_{\alpha} X^I \partial_{\beta} X^J \partial_{\gamma} \psi
\Bigr).
\end{eqnarray}
In order to analyze easily, we approximate the action up to the quadratic order in $A_{\mu}^{\,\,\,\alpha}$. As a result, the Chern-Simons term $-\frac{1}{6} E^{\mu \nu \lambda} 
\epsilon^{\alpha \beta \gamma} 
A_{\mu \alpha} A_{\nu \beta} A_{\lambda \gamma}$
is neglected.
From equation of motion of $A_{\mu}^{\,\,\,\alpha}$, we obtain
\begin{equation}
A_{\mu}^{\,\,\,\alpha}=- h^{\alpha \beta} \Pi_{\beta \mu}, \label{on-shell}
\end{equation}
where
$\Pi_{\alpha}^{\,\,\,\mu}= \partial_{\alpha} X^{\mu}
-\frac{i}{2} \bar{\psi} \Gamma^{\mu} \partial_{\alpha} \psi$,
$h_{\alpha \beta}=\partial_{\alpha} X^I \partial_{\beta} X_I$ and $h^{\alpha \beta}$ is an inverse. 
As a result, the quadratic order approximation in $A_{\mu}^{\,\,\,\alpha}$
corresponds to the
quadratic order approximation in $\partial_{\alpha} X^{\mu}$ and $\partial_{\alpha} \psi$. 
By substituting (\ref{on-shell}), the action reduces to
\begin{equation}
\tilde{S}_{cl}
=
\int d^3 \sigma 
\Bigl(
-\frac{1}{2}\frac{1}{\sqrt{-g}}h
+\frac{1}{2}\sqrt{-g}
(h^{\alpha \beta} \Pi_{\alpha}^{\,\,\, \mu} \Pi_{\beta \mu} + \Lambda)
+\frac{i}{4} \bar{\psi} \Gamma_{IJ} E^{\alpha \beta \gamma} \partial_{\alpha} X^I \partial_{\beta} X^J \partial_{\gamma} \psi
\Bigr),
\end{equation}
where $h$ is a determinant of $h_{\alpha \beta}$.
From equation of motion of $g$, we obtain
\begin{equation}
g=\frac{h}{h^{\alpha \beta} \Pi_{\alpha}^{\,\,\, \mu} \Pi_{\beta \mu} +\Lambda}.\end{equation}
If we substitute this relation, the action reduces to
\begin{equation}
\tilde{S}_{cl}
=
\int d^3 \sigma 
\sqrt{-h (h^{\alpha \beta} \Pi_{\alpha}^{\,\,\, \mu} \Pi_{\beta \mu} +\Lambda)}
+\frac{i}{4} \bar{\psi} \Gamma_{IJ} E^{\alpha \beta \gamma} \partial_{\alpha} X^I \partial_{\beta} X^J \partial_{\gamma} \psi.
\end{equation}
Up to the quadratic order in $\partial_{\alpha} X^{\mu}$ and $\partial_{\alpha} \psi$, we obtain
\begin{equation}
\tilde{S}_{cl}
=
\int d^3 \sigma 
\sqrt{-h} (1+ h^{\alpha \beta} \Pi_{\alpha}^{\,\,\, \mu} \Pi_{\beta \mu})
+\frac{i}{4} \bar{\psi} \Gamma_{IJ} E^{\alpha \beta \gamma} \partial_{\alpha} X^I \partial_{\beta} X^J \partial_{\gamma} \psi, \label{identical}
\end{equation}
when $\Lambda=1$.

Let us compare this action with the supermembrane action in M-theory given by \cite{BergshoeffSezginTownsend}
\begin{eqnarray}
S_{M2}&=&\int d^3 \sigma
\Bigl(
\sqrt{-G}
+\frac{i}{4} \epsilon^{\alpha \beta \gamma} \bar{\psi} \Gamma_{MN} \partial_{\alpha} \psi
(\Pi_{\beta}^{\,\,\, M} \Pi_{\gamma}^{\,\,\, N}
+\frac{i}{2}\Pi_{\beta}^{\,\,\, M} \bar{\psi} \Gamma^N \partial_{\gamma}\psi \n
&& \qquad \qquad \qquad \qquad \qquad \qquad \qquad -\frac{1}{12} \bar{\psi} \Gamma^M \partial_{\beta}\psi \bar{\psi} \Gamma^N \partial_{\gamma}\psi)
\Bigr), 
\end{eqnarray}
where $M, N=0, \cdots, 10$, $G_{\alpha \beta}= \Pi_{\alpha}^{\,\,\, M} \Pi_{\beta M}$ and $\Pi_{\alpha}^{\,\,\, M}= \partial_{\alpha} X^{M}
-\frac{i}{2} \bar{\psi} \Gamma^{M} \partial_{\alpha} \psi$.
$\psi$ is a $SO(1, 10)$ Majorana fermion.
In order to compare, we fix kappa symmetry by imposing the same condition (\ref{kappa}) and obtain 
\begin{eqnarray}
S_{M2}&=&\int d^3 \sigma
\Bigl(
\sqrt{-G}
+\frac{i}{4} \epsilon^{\alpha \beta \gamma} 
\bigl(
\bar{\psi} \Gamma_{\mu \nu} \partial_{\alpha} \psi
(\Pi_{\beta}^{\,\,\, \mu} \Pi_{\gamma}^{\,\,\, \nu}
+\frac{i}{2}\Pi_{\beta}^{\,\,\, \mu} \bar{\psi} \Gamma^{\nu} \partial_{\gamma}\psi
-\frac{1}{12} \bar{\psi} \Gamma^{\mu} \partial_{\beta}\psi \bar{\psi} \Gamma^{\nu} \partial_{\gamma}\psi) \n
&& \qquad \qquad \qquad \qquad \qquad  +\bar{\psi} \Gamma_{IJ} \partial_{\alpha} \psi \partial_{\beta} X^I \partial_{\gamma} X^J
\bigr)
\Bigr), 
\end{eqnarray}
where $G_{\alpha \beta}=h_{\alpha \beta} + \Pi_{\alpha}^{\,\,\, \mu} \Pi_{\beta \mu}$. If we take the same approximation, that is the approximation up to the quadratic order in $\partial_{\alpha} X^{\mu}$ and $\partial_{\alpha} \psi$, we obtain
\begin{equation}
\tilde{S}_{M2}
=
\int d^3 \sigma 
\sqrt{-h} (1+ h^{\alpha \beta} \Pi_{\alpha}^{\,\,\, \mu} \Pi_{\beta \mu})
+\frac{i}{4} \bar{\psi} \Gamma_{IJ} E^{\alpha \beta \gamma} \partial_{\alpha} X^I \partial_{\beta} X^J \partial_{\gamma} \psi.
\end{equation}
This action is identical with (\ref{identical}). As a result, we have shown that a supermembrane action in a semi-light-cone gauge (\ref{kappa}) can be written only with Nambu-Poisson bracket and the invariant symmetric bilinear form up to the quadratic order approximation in $\partial_{\alpha} X^{\mu}$ and $\partial_{\alpha} \psi$.

\section{Models of M-theory}
\setcounter{equation}{0}
In this section, we propose two 3-algebraic models of M-theory which are obtained by replacing Nambu-Poisson bracket in the action  (\ref{continuumaction}) with finite-dimensional 3-algebras' brackets as a second quantization.

If we replace Nambu-Poisson bracket in the action (\ref{continuumaction}) with a completely antisymmetric real 3-algebra's bracket \cite{kac, Class},
\begin{eqnarray}
&& \int d^3\sigma \sqrt{-g} \to \Bigl< \quad \Bigr> \n
&& 
\{ \varphi^a, \varphi^b, \varphi^c \}
\to [T^a, T^b, T^c], \label{classicallimit}
\end{eqnarray}
we obtain a second quantized model describing M-theory,
\begin{eqnarray}
S_{0}&=&
\Bigl<
-\frac{1}{12}[X^I, X^J, X^K]^2 
-\frac{1}{2}(A_{\mu a b} [T^a, T^b, X^{I}])^2 \n
&& \quad 
-\frac{1}{3} E^{\mu \nu \lambda} 
A_{\mu a b} A_{\nu c d} A_{\lambda e f} 
[T^a, T^c, T^d][T^b, T^e, T^f] \n
&& \quad 
-\frac{i}{2}\bar{\Psi} \Gamma^{\mu} A_{\mu a b} [T^a, T^b, \Psi] 
+\frac{i}{4}\bar{\psi}\Gamma_{IJ}[X^I, X^J, \psi] 
\Bigr>. \label{N=8model}
\end{eqnarray}
We have deleted the cosmological constant $\Lambda$, which corresponds to an operator ordering ambiguity, as usual as in the case of other matrix models \cite{BFSS, IKKT}.

This model can be obtained formally by a dimensional reduction of the $\mathcal{N}=8$ BLG model \cite{Lorentz1, Lorentz2, Lorentz3, Iso, Ghost-Free},
\begin{eqnarray}
S_{\mathcal{N}=8 BLG}&=&\int d^3x 
\Bigl<
-\frac{1}{12}[X^I, X^J, X^K]^2
-\frac{1}{2}(D_{\mu}X^{I})^2 
-E^{\mu \nu \lambda}
\bigl(\frac{1}{2} A_{\mu a b}\partial_{\nu} A_{\lambda c d} T^a [T^b, T^c, T^d]\n
&& \qquad \qquad
+\frac{1}{3} A_{\mu a b} A_{\nu c d} A_{\lambda e f} [T^a, T^c, T^d][T^b, T^e, T^f]\bigr) \n
&& \qquad \qquad
+\frac{i}{2}\bar{\Psi} \Gamma^{\mu} D_{\mu} \Psi
+\frac{i}{4}\bar{\psi}\Gamma_{IJ}[X^I, X^J, \psi]
\Bigr>. 
\end{eqnarray}
Therefore, the model (\ref{N=8model}) possesses 16 dynamical and 16 kinematical supersymmetries that form $\mathcal{N}=1$ space-time supersymmetry in eleven dimensions.

Next, the supermembrane action (\ref{continuumaction}) can be rewritten by using the triality of $SO(8)$ and the $SU(4) \times U(1)$ decomposition \cite{N=6BL, NishinoRajpoot, GustavssonRey} as
\begin{eqnarray}
S_{cl}
&=&
\int d^3 \sigma \sqrt{-g}
\Bigl(
-V
-A_{\mu b a}\{Z^A, T^a, T^b\}
A_{d c}^{\mu}\{Z_A, T^c, T^d\} \n
&&
\qquad \qquad +\frac{1}{3} E^{\mu \nu \lambda} A_{\mu b a} A_{\nu d c} A_{\lambda f e} \{T^a, T^c, T^d\} \{T^b, T^f, T^e\} \n
&&
\qquad \qquad +i\bar{\psi}^A \Gamma^{\mu} A_{\mu b a}\{\psi_A, T^a, T^b\}
+\frac{i}{2} E_{ABCD} \bar{\psi}^A \{Z^C, Z^D, \psi^B\}
-\frac{i}{2} E^{ABCD} Z_D \{\bar{\psi}_A, \psi_B,  Z_C\}
\n
&&
\qquad \qquad -i\bar{\psi}^A\{\psi_A, Z^B, Z_B\} 
+2i\bar{\psi}^A\{\psi_B, Z^B, Z_A\} 
\Bigr), \label{N=6matrixmodel}
\end{eqnarray}
where fields with a raised $A$ index transform in the $\bold{4}$ of SU(4), whereas those with lowered one transform in the $\bold{\bar{4}}$. $A_{\mu b a}$ ($\mu = 0, 1, 2$) is an anti-Hermitian gauge field, $Z^A$ and $Z_{A}$ are a complex scalar field and its complex conjugate, respectively. $\psi_{A}$ is a fermion field that satisfies
\begin{equation}
\Gamma^{012}\psi_{A}=-\psi_{A}, \label{SU(4)kappa}
\end{equation} 
and $\psi^{A}$ is its complex conjugate. 
$E^{\mu \nu \lambda}$ and $E^{ABCD}$ are Levi-Civita symbols in three dimensions and four dimensions, respectively. 
The potential terms are given by
\begin{eqnarray}
V &=& \frac{2}{3}\Upsilon^{CD}_B \Upsilon_{CD}^B \n
\Upsilon^{CD}_B &=& \{Z^C, Z^D, Z_B\}
-\frac{1}{2} \delta^C_B\{Z^E, Z^D, Z_E\}
+\frac{1}{2} \delta^D_B\{Z^E, Z^C, Z_E\}.
\end{eqnarray}

If we replace Nambu-Poisson bracket with a Hermitian 3-algebra's bracket \cite{LieOrigin, CherkisDotsenkoSaeman},
\begin{eqnarray}
&& \int d^3\sigma \sqrt{-g} \to \Bigl< \quad \Bigr> \n
&& 
\{ \varphi^a, \varphi^b, \varphi^c \}
\to [T^a, T^b ; \bar{T}^{\bar{c}}], \label{classicallimit2}
\end{eqnarray}
we obtain another second quantized model describing M-theory,
\begin{eqnarray}
S&=&
\Bigl<
-V
-A_{\mu \bar{b} a}[Z^A, T^a ; \bar{T}^{\bar{b}}]
\overline{A_{\bar{d} c}^{\mu}[Z_A, T^c ; \bar{T}^{\bar{d}}]}
+\frac{1}{3} E^{\mu \nu \lambda} A_{\mu \bar{b} a} A_{\nu \bar{d} c} A_{\lambda \bar{f} e} [T^a, T^c ; \bar{T}^{\bar{d}}] \overline{[T^b, T^f ; \bar{T}^{\bar{e}}]} \n
&&
+i\bar{\psi}^A \Gamma^{\mu} A_{\mu \bar{b} a}[\psi_A, T^a ; \bar{T}^{\bar{b}}]
+\frac{i}{2} E_{ABCD} \bar{\psi}^A [Z^C, Z^D ; \bar{\psi}^B]
-\frac{i}{2} E^{ABCD} \bar{Z}_D [\bar{\psi}_A, \psi_B ; \bar{Z}_C]
\n
&&
-i\bar{\psi}^A[\psi_A, Z^B ; \bar{Z}_B] 
+2i\bar{\psi}^A[\psi_B, Z^B ; \bar{Z}_A] 
\Bigr>, \label{N=6matrixmodel}
\end{eqnarray}
where the cosmological constant has been deleted for the same reason as before. The potential terms are given by
\begin{eqnarray}
V &=& \frac{2}{3}\Upsilon^{CD}_B \overline{\Upsilon}_{CD}^B \n
\Upsilon^{CD}_B &=& [Z^C, Z^D ; \bar{Z}_B]
-\frac{1}{2} \delta^C_B[Z^E, Z^D ; \bar{Z}_E]
+\frac{1}{2} \delta^D_B[Z^E, Z^C ; \bar{Z}_E].
\end{eqnarray}

This matrix model can be obtained formally by a dimensional reduction of the $\mathcal{N}=6$ BLG action, 
\begin{eqnarray}
S_{\mathcal{N}=6 BLG}&=&\int d^3x 
\Bigl<
-V
-D_{\mu} Z^A \overline{D^{\mu}Z_A} 
+E^{\mu \nu \lambda} 
\bigl(
\frac{1}{2} A_{\mu \bar{c} b}\partial_{\nu} A_{\lambda \bar{d} a} \bar{T}^{\bar{d}} [T^a, T^b ; \bar{T}^{\bar{c}}] \n
&& \qquad \quad
+\frac{1}{3} A_{\mu \bar{b} a} A_{\nu \bar{d} c} A_{\lambda \bar{f} e} [T^a, T^c ; \bar{T}^{\bar{d}}] \overline{[T^b, T^f ; \bar{T}^{\bar{e}}]}
\bigr) \n
&& \qquad \quad
-i\bar{\psi}^A \Gamma^{\mu} D_{\mu} \psi_A
+\frac{i}{2} E_{ABCD} \bar{\psi}^A [Z^C, Z^D ; \psi^B]
-\frac{i}{2} E^{ABCD} \bar{Z}_D [\bar{\psi}_A, \psi_B ; \bar{Z}_C] \n
&& \qquad \quad
-i\bar{\psi}^A[\psi_A, Z^B ; \bar{Z}_B] 
+2i\bar{\psi}^A[\psi_B, Z^B ; \bar{Z}_A] 
\Bigr>.
\end{eqnarray}
Therefore, the model (\ref{N=6matrixmodel}) has the explicit 12 dynamical supersymmetries inherited from the $\mathcal{N}=6$ BLG theory. Moreover, this model has implicit 4 dynamical supersymmetries when Chern-Simon level is one \cite{ABJM, GustavssonRey}. As a result, this model with Chern-Simon level one also possesses 16 dynamical and 16 kinematical supersymmetries that form $\mathcal{N}=1$ space-time supersymmetry in eleven dimensions.


\section{DLCQ and Reduction to BFSS Matrix Theory}
\setcounter{equation}{0}

It was shown that M-theory in DLCQ limit reduces to BFSS matrix theory with finite $n$ \cite{Susskind, Sen, Seiberg, Polchinski, textbook, BeckerBeckerSchwarz}. This fact is a strong criterion for a model of M-theory. In this section, we will take a specific Hermitian 3-algebra in the SU(4) model (\ref{N=6matrixmodel}) as an example and take DLCQ limit of it. As a result, we will obtain the BFSS matrix theory with finite $n$ as desired.

A Hermitian 3-algebra is constructed in \cite{N=6BL}:
\begin{eqnarray}
&&[T^a, T^b ; \bar{T}^{\bar{c}}]=\frac{2\pi}{k}(T^{a}T^{\dagger \bar{c}}T^{b}-T^{b}T^{\dagger \bar{c}}T^{a}) \n
&&\Bigl< \Bigr>=\Tr, \label{matrixrep}
\end{eqnarray}
where an integer $k$ represents Chern-Simons level and we choose $k=1$ in order to obtain 16 dynamical supersymmetries. $T^{a}$ span complete basis of $N \times N$ complex matrices and $T^{\dagger a}$ are their Hermitian conjugates. 
By using this, we obtain  
\begin{eqnarray}
S&=&\Tr
\Bigl(
-\frac{(2\pi)^2}{k^2} V-(Z^A A^R_{\mu}-A^L_{\mu} Z^A)(Z^A A^{R \mu}-A^{L \mu} Z^A)^{\dagger} 
-\frac{k}{2\pi}\frac{i}{3} E^{\mu \nu \lambda} (A^R_{\mu} A^R_{\nu} A^R_{\lambda}-A^L_{\mu} A^L_{\nu} A^L_{\lambda})\n
&&-\bar{\psi}^A \Gamma^{\mu} (\psi_A A^R_{\mu} -A^L_{\mu} \psi_A)  
+\frac{2\pi}{k}(i E_{ABCD} \bar{\psi}^A Z^{C} \psi^{\dagger B} Z^{D} 
-i E^{ABCD} Z^{\dagger}_{D} \bar{\psi^{\dagger}}_A Z^{\dagger}_{C}
\psi_B \n
&& -i \bar{\psi}^A \psi_A Z^{\dagger}_{B} Z^{B}
+i \bar{\psi}^A Z^{B} Z^{\dagger}_{B} \psi_A
+2i \bar{\psi}^A \psi_B Z^{\dagger}_{A} Z^{B}
-2i \bar{\psi}^A Z^{B} Z^{\dagger}_{A} \psi_B )
\Bigr), \label{ABJMrep}
\end{eqnarray}
where $A^R_{\mu} \equiv -\frac{k}{2\pi}iA_{\mu \bar{b} a}T^{\dagger \bar{b}}T^{a}$ and 
$A^L_{\mu} \equiv -\frac{k}{2\pi}iA_{\mu \bar{b} a}T^{a}T^{\dagger \bar{b}}$
are $N \times N$ Hermitian matrices.  
$V$ is given by
\begin{eqnarray}
V&=&+\frac{1}{3} Z^{\dagger}_{A} Z^{A} Z^{\dagger}_{B} Z^{B} Z^{\dagger}_{C} Z^{C}+\frac{1}{3} Z^{A} Z^{\dagger}_{A} Z^{B} Z^{\dagger}_{B} Z^{C} Z^{\dagger}_{C}+\frac{4}{3} Z^{\dagger}_{A} Z^{B} Z^{\dagger}_{C} Z^{A} Z^{\dagger}_{B} Z^{C} \n
&&-Z^{\dagger}_{A} Z^{A} Z^{\dagger}_{B} Z^{C} Z^{\dagger}_{C} Z^{B}
-Z^{A} Z^{\dagger}_{A} Z^{B} Z^{\dagger}_{C} Z^{C} Z^{\dagger}_{B}. 
\end{eqnarray}
This action can be obtained formally by a dimensional reduction of ABJM action \cite{ABJM, ABJ, PangWang}\footnote
{
The authors of \cite{HanadaMannelliMatsuo, IshikiShimasakiTsuchiya, KawaiShimasakiTsuchiya, IshikiShimasakiTsuchiya2} studied matrix models that can be obtained by a dimensional reduction of the ABJM and ABJ gauge theories on $S^3$. They showed that the models reproduce the original gauge theories on $S^3$ in planar limits.
}. Because ABJM action with level one is an effective action of N multiple supermembranes in the flat eleven-dimensional spacetime, the action should possess translational symmetry in the target space although it is not manifest. Therefore, the action (\ref{ABJMrep}) also should possess translational symmetry in the target space.

By redefining fields as
\begin{eqnarray}
Z^A&\to& \left(\frac{k}{2\pi}\right)^{\frac{1}{3}} Z^A \n
A^{\mu}&\to&  \left(\frac{2\pi}{k}\right)^{\frac{1}{3}} A^{\mu}\n
\psi^A&\to& \left(\frac{k}{2\pi}\right)^{\frac{1}{6}} \psi^A,
\end{eqnarray}
we obtain an action that is independent of Chern-Simons level:
\begin{eqnarray}
S&=&\Tr
\Bigl(
-V-(Z^A A^R_{\mu}-A^L_{\mu} Z^A)(Z^A A^{R \mu}-A^{L \mu} Z^A)^{\dagger} 
-\frac{i}{3} E^{\mu \nu \lambda} (A^R_{\mu} A^R_{\nu} A^R_{\lambda}-A^L_{\mu} A^L_{\nu} A^L_{\lambda})\n
&&-\bar{\psi}^A \Gamma^{\mu} (\psi_A A^R_{\mu} -A^L_{\mu} \psi_A)  
+i E_{ABCD} \bar{\psi}^A Z^{C} \psi^{\dagger B} Z^{D} 
-i E^{ABCD} Z^{\dagger}_{D} \bar{\psi^{\dagger}}_A Z^{\dagger}_{C}
\psi_B \n
&& -i \bar{\psi}^A \psi_A Z^{\dagger}_{B} Z^{B}
+i \bar{\psi}^A Z^{B} Z^{\dagger}_{B} \psi_A
+2i \bar{\psi}^A \psi_B Z^{\dagger}_{A} Z^{B}
-2i \bar{\psi}^A Z^{B} Z^{\dagger}_{A} \psi_B 
\Bigr),
\end{eqnarray}
as opposed to three-dimensional Chern-Simons actions.

If we rewrite the gauge fields in the action as $A^{L}_{\mu}=A_{\mu}+B_{\mu}$ and $A^{R}_{\mu}=A_{\mu}-B_{\mu}$, we obtain  
\begin{eqnarray}
S&=&\Tr
\Bigl(
-V
+([A_{\mu}, Z^A] + \{B_{\mu}, Z^A \})
([A^{\mu}, Z_A] - \{B^{\mu}, Z_A \})
+i E^{\mu \nu \lambda}(\frac{2}{3} B_{\mu} B_{\nu} B_{\lambda}
+2 A_{\mu} A_{\nu} B_{\lambda}) \n
&& +\bar{\psi}^A \Gamma^{\mu} ([A_{\mu}, \psi_A] + \{B_{\mu}, \psi_A \})
+i E_{ABCD} \bar{\psi}^A Z^{C} \psi^{\dagger B} Z^{D} 
-i E^{ABCD} Z^{\dagger}_{D} \bar{\psi^{\dagger}}_A Z^{\dagger}_{C}
\psi_B \n
&& -i \bar{\psi}^A \psi_A Z^{\dagger}_{B} Z^{B}
+i \bar{\psi}^A Z^{B} Z^{\dagger}_{B} \psi_A
+2i \bar{\psi}^A \psi_B Z^{\dagger}_{A} Z^{B}
-2i \bar{\psi}^A Z^{B} Z^{\dagger}_{A} \psi_B 
\Bigr), \label{action}
\end{eqnarray}
where $[\,\, , \,\,]$ and $\{\,\, , \,\,\}$ are the ordinary commutator and anticommutator, respectively. The $u(1)$ parts of $A^{\mu}$ decouple because $A^{\mu}$ appear only in commutators in the action. $B^{\mu}$ can be regarded as auxiliary fields, and thus $A^{\mu}$ correspond to matrices $X^{\mu}$ that represents three space-time coordinates in M-theory. Among $N \times N$ arbitrary complex matrices $Z^A$, we need to identify matrices $X^I$ ($I=3, \cdots 10$) representing the other space coordinates in M-theory, because the model possesses not $SO(8)$ but $SU(4) \times U(1)$ symmetry. Our identification is 
\begin{eqnarray}
Z^A &=& i X^{A+2} -X^{A+6}, \n
X^I &=& \hat{X}^I -i x^I \bold{1},
\end{eqnarray}
where $\hat{X}^I$ and $x^I$ are $su(N)$ Hermitian matrices and real scalars, respectively. This is analogous to the identification when we compactify ABJM action, which describes N M2 branes, and obtain the action of N D2 branes \cite{Lorentz0, ABJM, PangWang}. We will see that this identification works also in our case. We should note that while the $su(N)$ part is Hermitian, the $u(1)$ part is anti-Hermitian. That is, an eigen-value distribution of $X^{\mu}$, $Z^A$, and not $X^I$ determine the spacetime in the $SU(4)$ model. In order to define light-cone coordinates, we need to perform Wick rotation: $a^0 \to -i a^0$. After the Wick rotation, we obtain
\begin{equation}
A^{0}=\hat{A^{0}}-i a^{0} \bold{1}, 
\end{equation}
where $\hat{A^{0}}$ is a $su(N)$ Hermitian matrix.

Before taking DLCQ limit, we redefine fields as follows. First, by rescaling the eight matrices as
\begin{eqnarray}
X^I&=&\frac{1}{T}X^{'I}\n
A^{\mu}&=&A^{'\mu},
\end{eqnarray}
we adjust the scale of $X^I$ to that of $A^{\mu}$ and identify $A^{' \mu}$ with $X^{'\mu}$. $T$ is a real parameter. Next, we redefine fields so as to keep the scale of nine matrices:
\begin{eqnarray}
X^{'p}&=&X^{''p} \n
X^{'i}&=&X^{''i} \n
X^{'0}&=&\frac{1}{T}X^{''0} \n
X^{'10}&=&\frac{1}{T}X^{''10}, 
\end{eqnarray}
where $p=1,2$ and $i=3, \cdots, 9$. 
We also redefine auxiliary fields as 
\begin{equation}
B^{\mu} = \frac{1}{T^2} B^{'''\mu}.
\end{equation}

%

DLCQ limit of M-theory consists of a light-cone compactification, $x^- \approx x^- + 2 \pi R$, where $x^{\pm}=\frac{1}{\sqrt{2}}(x^{10} \pm x^0)$, and Lorentz boost in $x^{10}$ direction with an infinite momentum. 
In the following, we demonstrate DLCQ limit of the bosonic part of the model. One can obtain the same result in the fermionic part.  
 We define light-cone coordinates as
\begin{eqnarray}
X^{''0}&=&\frac{1}{\sqrt{2}}(X^+ - X^-) \n
X^{''10}&=&\frac{1}{\sqrt{2}}(X^+ + X^-). 
\end{eqnarray} 
We treat $B^{''' \mu}$ as scalars.  
A matrix compactification \cite{Taylor} on a circle with a radius R
imposes following conditions on $X^{-}$ and the other matrices $Y$ that represents $X^+$, $X^{''p}$, $X^{''i}$ and $B^{''' \mu}$:
\begin{eqnarray}
X^{-} -i (2\pi R) \bold{1} &=& U^{\dagger} X^{-} U \n
Y&=& U^{\dagger} Y U,
\label{periodic}
\end{eqnarray}
where $U$ is a unitary matrix. After the compactification, we cannot redefine fields freely.
A solution to (\ref{periodic}) is given by $U$, $X^{-} = \bar{X}^{-} + \tilde{X}^{-}$ and $Y=\tilde{Y}$, where
a unitary matrix $U$ is given by
\begin{equation}
U=
\left(
\begin{array}{cccc}
0&1&&0 \\
&\ddots&\ddots& \\
&&&1 \\
0&&&0
\end{array}
\right)
\otimes
\bold{1}_{n \times n},
\end{equation}
a background $\bar{X}^{-}$ is  
\begin{equation}
\bar{X}^{-}=
-i(T^3 \bar{x}^{-}) \bold{1} -i (2\pi R) \mbox{diag}(\cdots, s-1, s, s+1, \cdots) \otimes \bold{1}_{n \times n}, \label{bg}\end{equation}
and a fluctuation $\tilde{X}$ that represents $\tilde{X}^{-}$ and $\tilde{Y}$ is\begin{equation}
\tilde{X}=
\left(
\begin{array}{cccc}
\tilde{X}(0)&\tilde{X}(1)& \cdots & \\
\tilde{X}(-1)&\ddots&\ddots & \\
\vdots&\ddots&&\tilde{X}(1) \\
&&\tilde{X}(-1)&\tilde{X}(0)
\end{array}
\right). \label{fluctuation}
\end{equation}
Each $\tilde{X}(s)$ is a $n \times n$ matrix, where $s$ is an integer. That is, the (s, t)-th block is given by $\tilde{X}_{s,t} = \tilde{X}(s-t)$.

We make a Fourier transformation,
\begin{equation}
\tilde{X}(s)= \frac{1}{2 \pi \tilde{R}} \int^{2 \pi \tilde{R}}_0 d\tau X(\tau) e^{i s \frac{\tau}{\tilde{R}}},  \label{Fourier}
\end{equation}
where $X(\tau)$ is a $n \times n$ matrix in one-dimension and $R\tilde{R}=2\pi$. From (\ref{bg}), (\ref{fluctuation}) and (\ref{Fourier}), the following identities hold:
\begin{eqnarray}
&&
\sum_t \tilde{X}_{s,t} \tilde{X'}_{t, u}
=
\frac{1}{2 \pi \tilde{R}} \int^{2 \pi \tilde{R}}_0 d\tau \,
X(\tau) X'(\tau) e^{i(s-u) \frac{\tau}{\tilde{R}}} \n
&&
\tr(\sum_{s, t} \tilde{X}_{s,t} \tilde{X'}_{t,s})
=
V\frac{1}{2 \pi \tilde{R}} \int^{2 \pi \tilde{R}}_0 d\tau \, \tr(X(\tau)X'(\tau)) \n
&&
[\bar{X}^-, \tilde{X}]_{s, t} 
=
\frac{1}{2 \pi \tilde{R}} \int^{2 \pi \tilde{R}}_0 d\tau \,  \partial_{\tau} X(\tau) 
e^{i(s-t) \frac{\tau}{\tilde{R}}}, \label{id}
\end{eqnarray}
where $\tr$ is a trace over $n \times n$ matrices and $V=\sum_s 1$. We will use these identities later.

Next, let us boost the system in $x^{10}$ direction:
\begin{eqnarray}
X^+ &=& \frac{1}{T} X^{'''+} \n
\tilde{X}^- &=& T \tilde{X}^{'''-} \n
X^{''p}&=&X^{'''p} \n
X^{''i}&=&X^{'''i}.
\end{eqnarray}
IMF limit is achieved when $T \to \infty$.  
The second equation implies that $X^- \equiv -i(T^3 \bar{x}^{-}) \bold{1} +T X^{'''-}$,
where
$X^{'''-}=\bar{X}^{'''-} + \tilde{X}^{'''-} $
and $ \bar{X}^{'''-} =
-i (2\pi R') \mbox{diag}(\cdots, s-1, s, s+1, \cdots) \otimes \bold{1}_{n \times n}$. $R' \equiv \frac{1}{T} R$ goes to zero when $T \to \infty$.

To summarize, relations between the original fields and the fixed fields when $T \to \infty$ are 
\begin{eqnarray}
A^0 &=& \frac{1}{\sqrt{2}}(\frac{1}{T^2} X^{'''+} - X^{'''-})
+\frac{i}{\sqrt{2}}T^2 \bar{x}^{-} \bold{1} \n
A^p &=& X^{'''p} \n
X^i &=& \frac{1}{T}X^{'''i} \n
X^{10} &=& \frac{1}{\sqrt{2}}(\frac{1}{T^3} X^{'''+}+\frac{1}{T} X^{'''-})
-\frac{i}{\sqrt{2}}T\bar{x}^{-} \bold{1} \n
B^{\mu} &=& \frac{1}{T^2} B^{'''\mu}. \label{scaling}
\end{eqnarray}
By using these relations, equations of motion of the auxiliary fields $B^{\mu}$ are given by
\begin{eqnarray}
B^{''' 0}&=&\frac{i}{2 (\bar{x}^{''' -})^2} [X^{''' 1}, X^{''' 2}] + O(\frac{1}{T}) \n
B^{''' 1}&=&\frac{i }{2 \sqrt{2} (\bar{x}^{-})^2} [X^{''' 2}, X^{''' -}]
-\frac{i}{2 \bar{x}^{-}} [X^{''' 1}, X^{''' -}]  + O(\frac{1}{T}) \n
B^{''' 2}&=&-\frac{i}{2 \sqrt{2} (\bar{x}^{-})^2} [X^{''' 1}, X^{''' -}]
-\frac{i}{2 \bar{x}^{-}} [X^{''' 2}, X^{''' -}]  + O(\frac{1}{T}). 
\end{eqnarray}
If we substitute them and (\ref{scaling}) to the bosonic part of the action (\ref{action}), we obtain
\begin{eqnarray}
S_b&=& \frac{1}{T^2} \Tr 
\Bigl(
\frac{1}{4(\bar{x}^{-})^2}(-[X^{''' -}, X^{''' p}]^2 + [X^{''' p}, X^{''' q}]^2)
-\frac{1}{2}[X^{''' -}, X^{''' i}]^2
+[X^{''' p}, X^{''' i}]^2 \n
&& \qquad + (\bar{x}^{-})^2[X^{''' i}, X^{''' j}]^2
\Bigr) 
+ O(\frac{1}{T^3}).
\end{eqnarray}
Therefore, the bosonic part reduces to
\begin{eqnarray}
\hat{S}&=& \frac{1}{T^2} \Tr 
\Bigl(
\frac{1}{4(\bar{x}^{-})^2}(-[X^{''' -}, X^{''' p}]^2 + [X^{''' p}, X^{''' q}]^2)-\frac{1}{2}[X^{''' -}, X^{''' i}]^2
+[X^{''' p}, X^{''' i}]^2 \n
&& \qquad + (\bar{x}^{-})^2[X^{''' i}, X^{''' j}]^2
\Bigr) 
\end{eqnarray} 
in $T \to \infty$ limit.
By redefining
\begin{eqnarray}
\frac{1}{\sqrt{T}}\frac{1}{\sqrt{2\bar{x}^{-}}}X^{''' -}  &\to& X^{''' -} \n
\frac{1}{\sqrt{T}}\frac{1}{\sqrt{\bar{x}^{-}}}X^{''' p}  &\to& X^{''' p} \n
\frac{1}{\sqrt{T}}\sqrt{2\bar{x}^{-}}X^{''' i}  &\to& X^{''' i}, \end{eqnarray}
we obtain
\begin{equation}
\hat{S} = \Tr(-\frac{1}{2}[X^{''' -}, X^{''' I}]^2 +\frac{1}{4}[X^{''' I}, X^{''' J}]^2). \label{preBFSS}
\end{equation}
The background in $X^{''' -}$ is modified, where $\frac{1}{\sqrt{T}}R' \to R'$.
By using the identities (\ref{id}), we can rewrite (\ref{preBFSS}) and obtain the action of BFSS matrix theory with finite $n$,
\begin{equation}
\hat{S}=M \int^{\infty}_0 d\tau 
\tr (\frac{1}{2}(D_0 X^I)^2 + \frac{1}{4}[X^I, X^J]^2),
\end{equation}
after Wick rotation back: $\partial_{\tau} \to -i \partial_{\tau}$. We have used $\tilde{R'} = \infty$ because $R' \to 0$ when $T \to \infty$. In DLCQ limit of our model, we see that $X^-$ disappears and $X^+$ changes to $\tau$ as in the case of the light-cone gauge fixing of the membrane theory. 

Let us consider the case where our model is compactified on a torus $T^p$. If we take DLCQ limit of our model (\ref{action}) on $T^p$ in a similar way, and perform T-duality transformations along all of the torus directions, we obtain u(n) (p+1)-dimensional maximally supersymmetric Yang-Mills theory (SYM) on a dual torus $\tilde{T}^p$. The Yang-Mills coupling constant is given by $g_{YM}^2 = M_{11}^{3(2-p)} R_{10}^{3-p} \prod_{i=1}^p \frac{1}{R_i}$, where $M_{11}$ is the eleven-dimensional Planck mass, $R_{10}$ is a radius of $S^1$ in the 10-th direction, and  $R_i$ are sizes of cycles in $T^p$. We see that the theory is in a weak coupling region when $p \le 3$, while it is in a strong coupling region when $p \ge 4$, because $R_{10} \to 0$ in DLCQ limit.

Here we explain what we have done in this section in the brane language. Fundamental objects in M-theory are M2-branes, M5-branes and graviton states. Let us see what kind of states of these objects survive in DLCQ limit as in \cite{textbook, BeckerBeckerSchwarz}. These states reduce to objects in type IIA string theory because M-theory reduces to it. In DLCQ limit, energy of a state is given by $E=P_{10}+O(\frac{1}{p_{10}})$ because $R_{10} \to 0$ and a Kaluza-Klein (K-K) momentum $P_{10}=\frac{n}{R_{10}} \to \infty$. The leading contribution $P_{10}=n \frac{1}{g \sqrt{\alpha'}}$ is static energy of n D0-branes. This implies that a n-th K-K mode of an eleven-dimensional graviton survives and reduces to n D0-branes. The second term $O(\frac{1}{p_{10}})=O(R_{10})$ comes from energy of M2- and M5-branes wrapped around the $S^1$. Because their energy is $O(\frac{1}{p_{10}})$, only ground states of M2- and M5-branes that are completely wrapped around $S^1$ and $T^4$ survive, and they reduce to fundamental strings wrapped around one cycle of $T^p$ and D4-branes wrapped around four cycles of $T^p$, respectively. M5-branes cannot survive when $p \le 3$. Let us suppose that these M-branes have zero K-K momentum. Energy of them is given by $E=\sqrt{\bold{p}^2 + m^2}$, where $\bold{p}$ is a transverse momentum and $m$ is a mass of the M-branes. Because $E=O(m)=O(R_{10})$, $\bold{p}$ needs to be zero. This is a point of measure zero. Therefore, the M-branes need to have non-zero K-K momentum, and thus the fundamental strings need to end on the D0-branes, and the D4-branes need to form D0-D4 bound states. If we perform T-duality transformations along all of the torus directions, the D0-branes and all of the surviving fundamental strings are described by the u(n) (p+1)-dimensional SYM on $\tilde{T}^p$, and the D0-D4 bound state that consists of the n D0-branes and the N D4-branes becomes a D0-D4 bound state that consists of n D4-branes and N D0-branes when $p=4$. This bound state is described as a N BPS instanton in the u(n) (4+1)-dimensional SYM on $\tilde{T}^4$. As a result, the fundamental objects in M-theory on $T^p$ in DLCQ limit are described by the u(n) (p+1)-dimensional SYM on $\tilde{T}^p$. Because our model on $T^p$ in the limit reduces to this theory, we see that in DLCQ limit, our model describes M2-branes, M5-branes and graviton states. This fact suggests that our model includes correct physical degrees of freedom of M-theory.

Next, we discuss in the supergravity point of view. The eleven-dimensional supergravity compactified on $T^p$ is the (11-p)-dimensional maximal supergravity. Its global symmetry is summarized in the above table. 
\begin{table}
\begin{tabular}{|c||c|c|}
\hline
 & global symmetry in SUGRA & U-duality \\
\hline
M on $T^1$ & $SO(1,1,\bold{R})$ & - \\
\hline
M on $T^2$ & $SL(2, \bold{R}) \times SO(1,1,\bold{R})$ &  $SL(2, \bold{Z})$  \\
\hline
M on $T^3$ & $SL(2, \bold{R}) \times SL(3, \bold{R})$ & $SL(2, \bold{Z}) \times SL(3, \bold{Z})$  \\
\hline
M on $T^4$ & $SL(5, \bold{R})$ & $SL(5, \bold{Z})$ \\
\hline
\end{tabular}
\caption{Global symmetry in supergravity and U-duality in string theory when M-theory is compactified on tori.}
\label{table}
\end{table}
The origin of this symmetry is U-duality in string theory, which is obtained by quantizing the global symmetry, where $\bold{R}$ is replaced by $\bold{Z}$. The difference between the global symmetry of supergravities and U-duality is an accidental symmetry in low-energy effective theories. U-duality is also summarized in the table \cite{textbook, BeckerBeckerSchwarz}. Let us see that our model on $T^p$ in DLCQ limit, namely the (p+1)-dimensional SYM on $\tilde{T}^p$ can reproduce U-duality as in \cite{UdualSusskind, SethiSusskind, Rozali, BerkoozRozaliSeiberg}. In the $T^2$ case, our model reduces to the (2+1)-dimensional SYM on $\tilde{T}^2$ and it has a modular group of $\tilde{T}^2$, $SL(2, \bold{Z})$. This is identical with the U-duality group. In the $T^3$ case, the (3+1)-dimensional SYM on $\tilde{T}^3$ has not only a modular group of $\tilde{T}^3$, $SL(3, \bold{Z})$, but also $SL(2, \bold{Z})$ S-duality. The total group coincides with the U-duality group. In the $T^4$ case, because the (4+1)-dimensional SYM is not renormalizable and in a strong coupling region, we use the brane picture. This theory represents n D4-branes on $\tilde{T}^4$ in a strong coupling region, which is n M5-branes on $\tilde{T}^5$ in the M-theory point of view. This system has a modular group of $\tilde{T}^5$, $SL(5, \bold{Z})$, which coincides with the U-duality group. These facts suggest that our model can reproduce the eleven-dimensional supergravity.


\section{Conclusion and Discussion}
\setcounter{equation}{0}
In this paper, we have shown that a supermembrane action in an eleven-dimensional spacetime with a semi-light-cone gauge can be written only with Nambu-Poisson bracket and an invariant symmetric bilinear form up to the quadratic order approximation in $\partial_{\alpha} X^{\mu}$ and $\partial_{\alpha} \psi$ ($\alpha=0, 1, 2$ and $\mu=0, 1, 2$), and thus it has manifest VPD covariance even when the world-volume metric is flat. We have proposed two 3-algebraic models describing M-theory which are obtained as a second quantization of an action that is equivalent to the supermembrane action under the approximation. The second quantization is defined by replacing Nambu-Poisson bracket with finite-dimensional 3-algebras' ones.  The models possess space-time $\mathcal{N}=1$ supersymmetry in eleven dimensions, which consists of 16 kinematical and 16 dynamical supersymmetries. Although the models possess not full $SO(1,10)$ but $SO(1,2) \times SO(8)$ or $SO(1,2) \times SU(4) \times U(1)$ covariance, they possess the matrices $A^{\mu}$ and $X^{I}$ or $Z^A$ corresponding to the eleven coordinates in M-theory. $A^{\mu}$ are dynamical because our models possess the same structure as the supermembrane action with a semi-light-cone gauge, where only the kappa symmetry is fixed and bosonic space-time coordinates are not constrained. We have also shown that the $SU(4)$ model with the algebra (\ref{matrixrep}) reduces to BFSS matrix theory if we take DLCQ limit.

Here we assume the existence of a covariant matrix model of M-theory and discuss a relation to our models, although it seems currently impossible to construct it. The covariant model should possess eleven bosonic matrices and thirty-two fermionic matrices that represent eleven-dimensional $\mathcal{N}=1$ supercoordinates. Because physical degrees of freedom of bosons and fermions are eight and sixteen, respectively, the covariant model should possess a symmetry analogous to the kappa symmetry of the supermembrane action. If this symmetry is fixed, that is a semi-light-cone gauge is taken, the covariant model should reduce to one of our models as in the supermembrane case. In this gauge, bosonic matrices are not constrained as in our models, whereas SO(1,10) covariance is broken. From this point of view, the number of D0 branes in our models should not be fixed in contrast to BFSS matrix theory. We hope that our models are sufficient to calculate all physical observables in M-theory.

\section*{Note Added}
While we are in the final stage of writing the manuscript, a paper appeared \cite{Nastase} in which models of M-theory are proposed. We note that their models are different with our models because their models are three-dimensional field theories, while ours are zero-dimensional ones. For example, the relation between their and our models is similar to the relation between $\mathcal{N}=4$ super Yang-Mills theory and the IIB matrix model. They are different theories and give different physics.

%
%
%

\section*{Acknowledgements}
We would like to thank T. Matsuo, F. Sugino, T. Tada, A. Tsuchiya and especially H. Kawai for valuable discussions.


\vspace*{0cm}




\begin{thebibliography}{99}


\bibitem{BFSS}T. Banks, W. Fischler, S.H. Shenker, L. Susskind, ``M Theory As A Matrix Model: A Conjecture," Phys. Rev. {\bf D55} (1997) 5112, hep-th/9610043.



\bibitem{BLG1}J. Bagger, N. Lambert, ``Modeling Multiple M2's," Phys. Rev. {\bf D75}: 045020, 2007, hep-th/0611108.

\bibitem{Gustavsson}A. Gustavsson, ``Algebraic structures on parallel M2-branes," Nucl. Phys. {\bf B811}: 66, 2009, arXiv:0709.1260 [hep-th].

\bibitem{BLG2}J. Bagger, N. Lambert, ``Gauge Symmetry and Supersymmetry of Multiple M2-Branes," Phys. Rev. {\bf D77}: 065008, 2008, arXiv:0711.0955 [hep-th].




\bibitem{ABJM}O. Aharony, O. Bergman, D. L. Jafferis, J. Maldacena, ``N=6 superconformal Chern-Simons-matter theories, M2-branes and their gravity duals," JHEP {\bf 0810}: 091, 2008, arXiv:0806.1218 [hep-th]

\bibitem{N=6BL}J. Bagger, N. Lambert, ``Three-Algebras and N=6 Chern-Simons Gauge Theories," Phys. Rev. {\bf D79}: 025002, 2009, arXiv:0807.0163 [hep-th]







\bibitem{Iso}Y. Honma, S. Iso, Y. Sumitomo, S. Zhang, ``Janus field theories from multiple M2 branes," Phys. Rev. {\bf D78}: 025027, 2008, arXiv:0805.1895 [hep-th].

\bibitem{talk}Talk given at JPS 2008 Autumn Meeting, Yamagata, 20-23 Sep. 2008 `` Nambu bracket as a new gauge symmetry in M theory," Y. Matsuo


\bibitem{bosonicM}M. Sato, `` Covariant Formulation of M-Theory," Int. J. Mod. Phys. {\bf A24} (2009), 5019, arXiv:0902.1333 [hep-th]

\bibitem{LeePark1}K. Lee, J.-H. Park, ``Partonic description of a supersymmetric p-brane," arXiv:1001.4532 [hep-th]


\bibitem{BUTSURI}Y. Matsuo, ``New Formulation of M-theory and Nambu Bracket," BUTSURI {\bf 65} 3 March (2010)







\bibitem{deWHN}B. de Wit, J. Hoppe, H. Nicolai, ``On the Quantum Mechanics of Supermembranes," Nucl. Phys. {\bf B305}: 545, 1988.




\bibitem{IKKT} N. Ishibashi, H. Kawai, Y. Kitazawa, A. Tsuchiya, ``A Large-N Reduced Model as Superstring," Nucl. Phys. {\bf B498} (1997) 467, hep-th/9612115. 





\bibitem{Nambu}Y. Nambu, ``Generalized Hamiltonian dynamics," Phys. Rev. {\bf D7}: 2405, 1973.

\bibitem{Yoneya}H. Awata, M. Li, D. Minic, T. Yoneya, ``On the Quantization of Nambu Brackets," JHEP {\bf 0102} (2001) 013, hep-th/9906248.







\bibitem{Smolin1}L. Smolin, ``M theory as a matrix extension of Chern-Simons theory," Nucl. Phys. {\bf B591} (2000) 227, hep-th/0002009.
 
\bibitem{Smolin2}L. Smolin, ``The cubic matrix model and a duality between strings and loops," hep-th/0006137.

\bibitem{Azuma}T. Azuma, S. Iso, H. Kawai, Y. Ohwashi, ``Supermatrix Models," Nucl. Phys. {\bf B610} (2001) 251, hep-th/0102168.



\bibitem{LeePark2}K. Lee, J.-H. Park, ``Three-algebra for supermembrane and two-algebra for superstring," JHEP {\bf 0904} (2009) 012, arXiv:0902.2417 [hep-th]



\bibitem{Carlip}S. Carlip, ``Loop Calculations For The Green-Schwarz Superstring," Phys. Lett. {\bf B186} (1987) 141.

\bibitem{Kallosh}R.E. Kallosh, ``Quantization of Green-Schwarz Superstring," Phys. Lett. {\bf B195} (1987) 369.

\bibitem{KazamaYokoi}Y. Kazama, N. Yokoi, ``Superstring in the plane-wave background with RR flux as a conformal field theory," JHEP {\bf 0803} (2008) 057, arXiv:0801.1561 [hep-th]




\bibitem{MSUSY}T. Banks, N. Seiberg, S. Shenker, ``Branes from Matrices," Nucl. Phys. {\bf B490}: 91, 1997, hep-th/9612157



\bibitem{M5fromM2}P-M. Ho, Y. Matsuo, ``M5 from M2," JHEP {\bf 0806} (2008) 105, arXiv:0804.3629 [hep-th]

\bibitem{M5withC}P-M. Ho, Y. Imamura, Y. Matsuo, S. Shiba, ``M5-brane in three-form flux and multiple M2-branes," JHEP {\bf 0808} (2008) 014, arXiv:0805.2898 [hep-th]




\bibitem{BergshoeffSezginTownsend}E. Bergshoeff, E. Sezgin, P.K. Townsend, ``Supermembranes and Eleven-Dimensional Supergravity," Phys. Lett. {\bf B189} (1987) 75.





\bibitem{kac}P.-M. Ho, Y. Matsuo, S. Shiba, ``Lorentzian Lie (3-)algebra and toroidal compactification of M/string theory," arXiv:0901.2003 [hep-th]

\bibitem{Class}P. de Medeiros, J. Figueroa-O'Farrill, E. Mendez-Escobar, P. Ritter, ``Metric 3-Lie algebras for unitary Bagger-Lambert theories," JHEP {\bf 0904}: 037, 2009, arXiv:0902.4674 [hep-th]






\bibitem{Lorentz1}J. Gomis, G. Milanesi, J. G. Russo, ``Bagger-Lambert Theory for General Lie Algebras," JHEP {\bf 0806}: 075, 2008, arXiv:0805.1012 [hep-th].

\bibitem{Lorentz2}S. Benvenuti, D. Rodriguez-Gomez, E. Tonni, H. Verlinde, ``N=8 superconformal gauge theories and M2 branes," JHEP {\bf 0901}: 078, 2009, arXiv:0805.1087 [hep-th].

\bibitem{Lorentz3}P.-M. Ho, Y. Imamura, Y. Matsuo, ``M2 to D2 revisited," JHEP {\bf 0807}: 003, 2008, arXiv:0805.1202 [hep-th].

\bibitem{Ghost-Free}M. A. Bandres, A. E. Lipstein, J. H. Schwarz, ``Ghost-Free Superconformal Action for Multiple M2-Branes," JHEP {\bf 0807}: 117, 2008, arXiv:0806.0054 [hep-th]

 






\bibitem{NishinoRajpoot}H. Nishino, S. Rajpoot, ``Triality and Bagger-Lambert Theory," Phys. Lett. {\bf B671}, (2009) 415 arXiv:0901.1173 [hep-th].

\bibitem{GustavssonRey}A. Gustavsson, S-J. Rey, ``Enhanced N=8 Supersymmetry of ABJM Theory on R(8) and R(8)/Z(2)," arXiv:0906.3568 [hep-th].







\bibitem{LieOrigin}P. de Medeiros, J. Figueroa-O'Farrill, E. Me'ndez-Escobar, P. Ritter, ``On the Lie-algebraic origin of metric 3-algebras," Commun. Math. Phys. {\bf 290} (2009) 871, arXiv:0809.1086 [hep-th]

\bibitem{CherkisDotsenkoSaeman}S. A. Cherkis, V. Dotsenko, C. Saeman, ``On Superspace Actions for Multiple M2-Branes, Metric 3-Algebras and their Classification," Phys. Rev. {\bf D79} (2009) 086002, arXiv:0812.3127 [hep-th]







\bibitem{Susskind}L. Susskind, ``Another Conjecture about M(atrix) Theory," hep-th/9704080
\bibitem{Sen}A. Sen, ``D0 Branes on $T^n$ and Matrix Theory,"  Adv. Theor. Math. Phys. {\bf 2} 51 (1998), hep-th/9709220
\bibitem{Seiberg}N. Seiberg, ``Why is the Matrix Model Correct?," Phys. Rev. Lett. {\bf 79} 3577 (1997), hep-th/9710009
\bibitem{Polchinski}J. Polchinski, ``M-Theory and the Light Cone," Prog. Theor. Phys. Suppl. {\bf 134} (1999) 158, hep-th/9903165
\bibitem{textbook}J. Polchinski, ``String Theory Vol. 2: Superstring Theory and Beyond" Cambridge University Press, Cambridge, UK, 1998
\bibitem{BeckerBeckerSchwarz}K. Becker, M. Becker, J. H. Schwarz, ``String Theory and M-theory," Cambridge University Press, Cambridge, UK, 2007



\bibitem{ABJ}O. Aharony, O. Bergman, D. L. Jafferis, ``Fractional M2-branes," JHEP {\bf 0811} (2008) 043, arXiv:0807.4924 [hep-th]

\bibitem{PangWang}Y. Pang, T. Wang, ``From N M2's to N D2's," Phys. Rev. {\bf D78} (2008) 125007, arXiv:0807.1444 [hep-th]




\bibitem{HanadaMannelliMatsuo}M. Hanada, L. Mannelli, Y. Matsuo, ``Large-N reduced models of supersymmetric quiver, Chern-Simons gauge theories and ABJM," arXiv:0907.4937 [hep-th]

\bibitem{IshikiShimasakiTsuchiya}G. Ishiki, S. Shimasaki, A. Tsuchiya, ``Large N reduction for Chern-Simons theory on $S^3$," Phys. Rev. {\bf D80} (2009) 086004, arXiv:0908.1711 [hep-th]

\bibitem{KawaiShimasakiTsuchiya}H. Kawai, S. Shimasaki, A. Tsuchiya, ``Large N reduction on group manifolds," arXiv:0912.1456 [hep-th]

\bibitem{IshikiShimasakiTsuchiya2}G. Ishiki, S. Shimasaki, A. Tsuchiya, ``A Novel Large-N Reduction on $S^3$: Demonstration in Chern-Simons Theory," arXiv:1001.4917 [hep-th]




\bibitem{Lorentz0}S. Mukhi, C. Papageorgakis, ``M2 to D2," JHEP {\bf 0805}: 085, 2008, arXiv:0803.3218 [hep-th].





\bibitem{Taylor}W. Taylor, ``D-brane field theory on compact spaces," Phys. Lett. {\bf B394} (1997) 283, hep-th/9611042



\bibitem{UdualSusskind}L. Susskind, ``T Duality in M(atrix) Theory and S Duality in Field Theory," hep-th/9611164

\bibitem{SethiSusskind}S. Sethi, L. Susskind, ``Rotational Invariance in the M(atrix) Formulation of Type IIB Theory," Phys. Lett. {\bf B400} (1997)  265, hep-th/9702101

\bibitem{Rozali}M. Rozali, ``Matrix Theory and U-Duality in Seven Dimensions," Phys. Lett. {\bf B400} (1997) 260, hep-th/9702136

\bibitem{BerkoozRozaliSeiberg}M. Berkooz, M. Rozali, N. Seiberg, ``Matrix Description of M-theory on $T^4$ and $T^5$," Phys. Lett. {\bf B408} (1997) 105, hep-th/9704089



\bibitem{Nastase}A. Mohammed, J. Murugan, H. Nastase, ``Looking for a Matrix model of ABJM," arXiv:1003.2599



























\end{thebibliography}
\end{document}